\documentclass[useAMS]{mn2e}
\usepackage{psfig}

\usepackage{times}
\usepackage{amssymb,amsmath}

\def\lsim{ \lower .75ex\hbox{$\sim$} \llap{\raise .27ex \hbox{$<$}} }
\def\gsim{ \lower .75ex \hbox{$\sim$} \llap{\raise .27ex \hbox{$>$}} }

\title[Spine-sheath BL Lac jets and IceCube neutrinos] 
{High-energy cosmic neutrinos from spine-sheath BL Lac jets}

\author[Tavecchio \& Ghisellini]
{F. Tavecchio$^1$\thanks{E--mail: fabrizio.tavecchio@brera.inaf.it} and
G. Ghisellini$^1$\\
$^1$INAF -- Osservatorio Astronomico di Brera, via E. Bianchi 46, I--23807
Merate, Italy\\
}

\begin{document}

% \date{Accepted 1988 December 15. Received 1988 December 14; 
% in original form 1988 October 11}

%\pagerange{\pageref{firstpage}--\pageref{lastpage}} \pubyear{2007}

\maketitle

\begin{abstract} 
We recently proposed that structured (spine-sheath) jets associated to BL Lac objects offer a suitable  
environment  for the production of the extragalactic high--energy ($E>100$ TeV) neutrino recently revealed by IceCube. Our previous analysis was limited to low--power BL Lac objects. We extend our preliminary study to the entire BL Lac population. We assume that the power of cosmic rays as well as 
the radiative luminosity of the sheath depend linearly on the the jet power.
In turn, we assume that the latter is well traced by the $\gamma$--ray luminosity.
We exploit the BL Lac $\gamma$--ray luminosity function and its cosmic evolution as recently 
inferred from {\it Fermi}--LAT data to derive the expected neutrino cumulative intensity 
from the entire BL Lac population. 
When considering only the low--power BL Lacs, a large cosmic ray power for each source
is required to account for the neutrino flux. 
Instead, if BL Lacs of all powers produce neutrinos, the power demand decreases,
and the required cosmic ray power becomes of the same order of the radiative jet power.
We also discuss the prospects for the direct association of IceCube events with BL Lacs, 
providing an estimate of the expected counts for the most promising sources.
\end{abstract}


\section{Introduction}

The detection of high-energy neutrinos by the IceCube observatory at 
the South Pole (Aartsen et al. 2013a, 2014) opened a new window for the study 
of the energetic astrophysical phenomena.
The discovery has triggered a wealth of studies devoted to the identification of the
possible sources (e.g., Anchordoqui et al. 2014 and Murase 2014 for recent reviews). 
The data are consistent with a flavor ratio  $\nu_{\rm e}:\nu_{\mu}:\nu_{\tau}=1:1:1$ 
and the flux level is close to the so--called Waxman--Bahcall limit (Waxman \& Bahcall 1999), 
valid if neutrinos  are produced by ultra--high energy cosmic rays (UHECR; $E>10^{19}$ eV) through 
pion--producing hadronic interaction before leaving their -- optically thin -- sources. 
However, the energies of the neutrinos ($E_{\nu}<$ few PeV) indicate that they are associated 
to cosmic rays with energies much below the UHECR regime, $E\lesssim10^{17}$ eV. 
The substantial isotropy of the flux (with only a non significant small excess in the 
direction of the galactic center) is consistent with an extragalactic origin, although 
a sizable contribution from galactic sources cannot be ruled out (e.g. Ahlers \& Murase 2014).  
Possible extragalactic astrophysical sources include propagating comic rays, star-forming and starburst galaxies, galaxy clusters, $\gamma$--ray burst and active galactic nuclei (AGN).

Among AGN, blazars, characterized by the presence of a relativistic jet of plasma moving 
toward the observer (e.g., Urry \& Padovani 1995), have been widely discussed in the past 
as candidates cosmic ray (CR) accelerators (e.g., Biermann \& Strittmatter 1987; see 
Kotera \& Olinto 2011 for a review) and thus potential neutrino emitters (e.g., Atoyan \& Dermer 2003, Becker 2008).  Murase et al. (2014) and Dermer et al. (2014) revisited the possibility -- already discussed in the past, e.g., 
Atoyan and Dermer (2003) -- that the observed neutrinos are produced in the jet of blazars 
through photo--pion reactions involving high energy CR and soft photons ($p+\gamma \to X +\pi$), 
followed by the prompt charged pion decay ($\pi ^{\pm}\to \mu^{\pm}+\nu_{\mu}\to e^{\pm} + 2\nu_{\mu} +\nu_{\rm e}$; hereafter we do not distinguish among $\nu$ and $\bar{\nu}$). 
Their analysis -- based on the simplest, one--zone, framework -- led to the conclusion 
that both the flux level and the spectral shape inferred by the IceCube data are difficult 
to reproduce by this scenario. 
In particular they predicted a rapid decline of the emission below 1 PeV. 
In their framework -- in which the CR luminosity is assumed to be proportional to the 
electromagnetic output -- it is naturally expected that the neutrino cumulative flux is 
dominated by the most luminous and powerful blazars, i.e. the flat spectrum radio quasars (FSRQ), 
which are also the sources characterized by the most rich radiative environment (required to have efficient photo-meson reactions). 
BL Lac objects, the low--power blazars defined as those to display faint or even absent 
optical broad emission lines, would provide only a minor contribution.

As noted, the Murase et al. (2014) analysis relies on the simplest scenario for blazars, 
assuming in particular that their jets are characterized by a well localized emission region 
(hence the definition of one--zone models) with a well defined speed. 
In a previous paper (Tavecchio, Ghisellini \& Guetta 2014, hereafter Paper I), we reconsidered 
this issue showing that, under the assumption that the jet presents a velocity structure, i.e. 
the flow is composed by a fast spine surrounded by a slower sheath (or layer), the neutrino output 
from the weak BL Lac objects (the so--called Highly peaked BL Lacs, HBL) is boosted and could match the observations. 
The  proposal for the existence of a velocity structure of the jet have been advanced  as a possible 
solution for the the so--called ``Doppler crisis" for TeV BL Lacs (e.g. Georganopoulos \& Kazanas 2003, 
Ghisellini, Tavecchio, \& Chiaberge 2005) and to unify the BL Lacs and radiogalaxy populations 
(e.g. Chiaberge et al. 2000, Meyer et al. 2011, Sbarrato et al. 2014). 
Direct  radio VLBI imaging of both radiogalaxies (e.g. Nagai et al. 2014, M{\"u}ller et al. 2014) and BL Lac 
(e.g., Giroletti et al. 2004, Piner \& Edwards 2014) jets, often showing a ``limb brightening" 
transverse structure, provides a convincing observational support to this idea, also corroborated by numerical simulations (e.g. McKinney 2006, Rossi et al. 2008). 
The reason behind the possibility to increase the neutrino (and inverse Compton $\gamma$-ray) production 
efficiency in such a spine--layer structure stems from the fact that for  particles flowing in 
the spine the radiation field produced in the layer appears to be amplified because of the 
relative motion between the two structures (e.g. Tavecchio \& Ghisellini 2008). 
In this conditions, the density of the soft photons in the spine rest frame -- determining the 
proton cooling rate and hence the neutrino luminosity -- can easily exceed that of the locally 
produced synchrotron ones, the only component taken in consideration in the one--zone modeling of 
Murase et al. (2014) in the case of BL Lacs (for FSRQ, instead, the photon field is thought to be 
dominated by the radiation coming from the external environment).

In Paper I we considered only the weakest BL Lac sources -- similar to the prototypical TeV 
blazar Mkn 421 --  for which the arguments supporting the existence of the jet structure are the most compelling. 
Interestingly, a hint of an actual association between the IceCube events and some 
low--power TeV emitting BL Lac (among which the aforementioned Mkn 421) has been found 
by Padovani \& Resconi (2014), although their result is not confirmed by  a 
sophisticated analysis of the IceCube collaboration (Aartsen et al. 2013b, 2014).  

There are hints suggesting that a velocity structure could be a universal characteristics of all BL Lac jets. 
This idea is supported by the modeling of the radio-galaxy emission through the spine-layer model 
(Tavecchio \& Ghisellini 2008, 2014), which suggests that these jets are typically more 
powerful than those associated to the weakest BL Lac (rather, they resembles the BL Lacs of the intermediate, 
IBL, or low--synchrotron peak, LBL, category). 
These arguments motivated us to extend our previous work presented in Paper I, considering the possibility 
that the entire BL Lac population is a source of high--energy neutrinos. 
To this aim we have to refine the simple description of the cosmic evolution we adopted in Paper I 
with a more complex, luminosity--dependent, evolution of the BL Lac luminosity function. 
We describe our neutrino emission model and the assumed cosmic evolution of BL Lacs in \S 2. 
We report the results in \S 3, in which also we present a list of the most probable candidates 
expected to be associated with the IceCube events. 
In \S 4 we conclude with a discussion.

Throughout the paper, the following cosmological  parameters are assumed: 
$H_0=70$ km s$^{-1}$ Mpc$^{-1}$, $\Omega_{\rm M}=0.3$, $\Omega_{\Lambda}=0.7$. 
We  use the notation $Q=Q_X \, 10^X $ in cgs units.

\section{Setting the stage}

\subsection{Neutrino emission}

We calculate the neutrino emission from a single BL Lac following the scheme 
already adopted and  described in Paper I. 
Here we just recall its basic features. 

We assume a two--flow jet structure, with a fast spine (with bulk Lorentz factor 
$\Gamma_{\rm s}$) with cross sectional radius $R$ surrounded by the slower and 
thin layer (with $\Gamma_{\rm l}<\Gamma_{\rm s}$). The corresponding Doppler factors, 
denoting with $\theta_{\rm v}$ the observing angle, are
$\delta_{\rm l,s}=[\Gamma_{\rm l,s}(1-\beta_{\rm l,s}\cos \theta_{\rm v})]^{-1}$. 

We further assume that the spine carries a population of high-energy CR (protons, for simplicity), 
whose luminosity in the spine frame (for which we use primed symbols) is parametrized by a 
cut--offed power law distribution in energy:
\begin{equation}
L^ {\prime}_{\rm p}(E_{\rm p}^ {\prime})=k_{\rm p} E_{\rm p}^{\prime \, -n}
\exp\left( -\frac{E_{\rm p}^ {\prime}}{E_{\rm cut}^ {\prime}}\right) \;\;\; 
E_{\rm p}^ {\prime}>E_{\rm min}^ {\prime}
\end{equation} 
with total (spine frame) luminosity 
$L_{\rm p}^ {\prime}=\int L_{\rm p}^ {\prime}(E_{\rm p}^ {\prime}) dE_{\rm p}^ {\prime}$.  
The cooling rate $t^{\prime \, -1}_{p\gamma}(E_{\rm p}^{\prime})$  of protons with energy 
$E_{\rm p}^{\prime}$ through the photo--meson reaction with a target radiation field with 
numerical density $n_{\rm t}^{\prime}(\epsilon)$ is given by (Atoyan \& Dermer 2003, 
see also Dermer \& Menon 2009):
\begin{equation}
t^{\prime \, -1}_{p\gamma}(E_{\rm p}^{\prime})=c \int_{\epsilon_{\rm th}}^{\infty} d\epsilon 
\frac{n_{\rm t}^ {\prime}(\epsilon)}{2\gamma_{\rm p}^{\prime}\epsilon^2} 
\int_{\epsilon_{\rm th}}^{2\epsilon\gamma_{\rm p}^{\prime}} d{\bar\epsilon}\, 
\sigma_{p\gamma}({\bar\epsilon})\, K_{p\gamma}({\bar\epsilon}) \,  {\bar\epsilon},
\label{tpg}
\end{equation}
where $\gamma _{\rm p}^{\prime}=E_{\rm p}^{\prime}/m_{\rm p}c^2$, $\sigma_{p\gamma}(\epsilon)$ 
is the photo--pion cross section, $K_{p\gamma}(\epsilon)$ the inelasticity and $\epsilon_{\rm th}$ 
is the threshold energy of the process. 
The photo--meson production efficiency is measured by the factor $f_{p\gamma}$, defined as 
the ratio between the timescales of the competing adiabatic and photo-meson  losses:
\begin{equation}
f_{p\gamma}(E_{\rm p}^ {\prime})=\frac{t^ {\prime}_{\rm ad}}{t^{\prime}_{p\gamma}(E_{\rm p}^{\prime})}.
\end{equation}
where $t^ {\prime}_{\rm ad}\approx R/c$.

The neutrino luminosity in the spine frame can thus  be calculated as:
\begin{equation}
E_{\nu}^{\prime} L^{\prime}_{\nu}({E_{\nu}^{\prime}}) \simeq \frac{3}{8} 
\min [1,f_{p\gamma}(E_{\rm p}^ {\prime})] \, E_{\rm p}^{\prime} L_{\rm p}^{\prime}({E_{\rm p}^{\prime}}),
\end{equation}
where $E_{\nu}^{\prime}=0.05\, E_{\rm p}^{\prime}$. 
The factor $3/8$ takes into account the fraction of the energy going into $\nu$ and $\bar{\nu}$ (of all flavors). 

The {\it observed} luminosity is derived taken into account the relativistic boosting, 
parametrized by the relativistic Doppler factor 
$\delta_{\rm s}$: $ E_{\nu} L_{\nu}({E_{\nu}}) = E_{\nu}^{\prime} L^{\prime}_{\nu}({E_{\nu}^{\prime}})\, 
\delta_{\rm s}^4$ and $E_{\nu}=\delta_{\rm s}E_{\nu}^{\prime}$.

We assume that the dominant population of soft photons -- 
specifying $n^{\prime}_{\rm t}(\epsilon)$ in Eq. (\ref{tpg}) -- is provided by the boosted layer radiation 
(we show in Paper I that the internally produced synchrotron photons provide a negligible contribution). 
The spectrum of this component is modeled as a broken power law $L(\epsilon_{\rm l})$ 
with indices $\alpha_{1,2}$ and (observer frame, unprimed symbols) SED peak energy $\epsilon_{\rm o}$. 
The layer luminosity is parametrized by the total (integrated) 
luminosity -- in the observer frame --  $L_{\rm l}$. 

As in Paper I we neglect the anisotropy of the layer radiation field in the spine frame (Dermer 1995).
We also neglect the high-energy photons  produced in the neutral pions decay $\pi^0\to \gamma \gamma$.

\subsection{Model parameters and scaling laws}

Summarizing, our model is specified by the following parameters: the jet radius, $R$, the spine and 
layer Lorentz factors $\Gamma_{\rm s}$ and $\Gamma_{\rm l}$, the observed layer radiative luminosity $L_{\rm l}$, the peak $\epsilon_{\rm o}$ of its energy distribution [in  $\epsilon L(\epsilon)$];
the spectral slopes $\alpha_1$ and $\alpha_2$  
of $L_{\rm l}(\epsilon)$ 
%[where $L_{\rm l}(\epsilon) \propto \epsilon^{-\alpha}]$
, the spine comoving 
CR luminosity $L^{\prime}_p$, the CR power law index $n$, the minimum and the cut--off energy 
$E^{\prime}_{\rm min}$, $E^{\prime}_{\rm cut}$. 

For definiteness we fix the structural jet parameters of the entire BL Lac 
population to the values adopted in Paper I. 
Specifically we assume  $\Gamma_{\rm s}=15$, $\Gamma_{\rm l}=2$, a jet radius $R=10^{15}$ cm and  
a layer spectrum with slopes $\alpha_{1}=0.5$ and $\alpha_{1}=1.5$. 

% ---------------------------------------------------
\begin{figure}
%\hskip -2 cm
%\hspace*{-0.3 truecm}
\vspace*{-1.5 truecm}
\hspace*{-0.8 truecm}
%\vskip -0.3 cm
\psfig{file=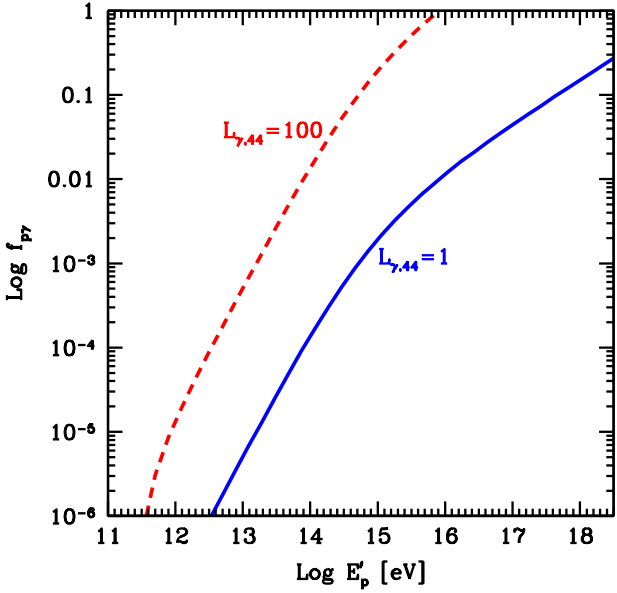,height=10.5cm,width=10.5cm}
\vspace{-1.1 cm}
\caption{Photopion production efficiency, $f_{p\gamma}$, as a function of the energy (in the jet spine frame) for 
protons carried by the spine scattering off the layer radiation field, for a BL Lac 
$\gamma$--ray luminosity $L_{\gamma}=10^{44}$ erg s$^{-1}$ (solid blue line) and 
$L_{\gamma}=10^{46}$ erg s$^{-1}$ (dashed red line).
}
\label{diffuse}
\end{figure}
% ---------------------------------------------------

In Paper I we considered two possible realizations of the model, characterized by two different values 
of the layer peak energy, $\epsilon_{\rm o}$. 
Through the threshold condition, $E^{\prime}_{\rm p}\epsilon^{\prime}_{\rm t}> m_{\pi}m_{\rm p}c^4$, 
$\epsilon_{\rm o}$ affects the possible values of the minimum CR energy. 
In fact, increasingly larger values of $\epsilon_{\rm o}$ allow for lower values of $E^{\prime}_{\rm p,min}$ 
and thus lower energies of the produced neutrinos. 
On the other hand, due to the steep CR distribution, decreasing $E^{\prime}_{\rm p,min}$ leads to increase the CR power required to produce a given neutrino output. In Paper I we show that to satisfactorily reproduce the low energy data points around 100 TeV we have 
to assume a layer emission peaking in the UV band, $\epsilon_{\rm o}\approx 400$ eV. 
In the following we adopt this value for our reference model.

For both the CR luminosity $L^{\prime}_{\rm p}$ and the layer radiative
luminosity $L_{\rm l}$ we assume, 
as a physically--motivated working hypothesis, a linear dependence on the jet power, 
$P_{\rm jet}$ --- i.e. constant efficiencies. 
In turn, we assume that $P_{\rm jet}$ is well traced by the observed $\gamma$--ray $0.1-100$ GeV 
luminosity, $L_{\gamma}$, as supported by the modeling of blazar SED (e.g. Ghisellini et al. 2014). 
We normalize the values of $L^{\prime}_{\rm p}$ and $L_{\rm l}$ to the values corresponding to 
the weakest sources, corresponding to $L_{\gamma}=10^{44}$ erg s$^{-1}$.
Therefore, we assume:
\begin{equation}
L^{\prime}_{\rm p}=L^{\prime}_{\rm p,o} \, \frac{L_{\gamma}}{10^{44} {\rm erg}\,\, {\rm s}^{-1}}; \; \;\;
L_{\rm l}=L_{\rm l,o} \, \frac{L_{\gamma}}{10^{44} {\rm erg}\,\, {\rm s}^{-1}}.
\end{equation}
For the layer we adopt the value used in  Paper I, $L_{\rm t,o}=2\times 10^{44}$ erg s$^{-1}$, 
while $L^{\prime}_{\rm p,o}$ is left as a free parameter, together with the other parameters 
specifying the CR distribution.
The value of $f_{p\gamma}(E^{\prime}_{\rm p})$ for the assumed set of parameters is shown in Fig. 1. Note the large efficiency ($f_{\rm p\gamma}>0.1$) characterizing the most powerful sources for proton energies corresponding the neutrinos detected by IceCube.

Given the assumed linear scaling of CR and layer luminosities with the jet power, 
the neutrino luminosity --- proportional to their product --- will scale as 
$L_{\nu}\propto L^{\prime}_{\rm p} L_{\rm t}\propto L^2_{\gamma}$. 
Alternatively, this can be expressed by the fact that the efficiency of the neutrino production, 
$\eta _{\nu}\equiv L_{\nu}/P_{\rm jet}$, increases with the jet power (and thus with 
the $\gamma$--ray luminosity), $\eta _{\nu} \propto L_{\gamma}$. 
We will see that this fact implies that, despite the cosmic density of sources decreases with 
their $\gamma$--ray luminosity (i.e. a decreasing luminosity function), the cumulative 
cosmic neutrino output is dominated by the most powerful --- but rare --- sources.

\subsection{Diffuse intensity}

The cumulative diffuse neutrino intensity deriving from the entire population of BL Lacs, is evaluated as:
\begin{equation}
E_{\nu}I(E_{\nu})= \frac{c E_{\nu}^2 }{4\pi H_o} \int_{0}^{z_{\rm max}} \int_{L_{\gamma,1}}^{L_{\gamma,2}} \frac{j[L_{\gamma},E_{\nu}(1+z),z]}{\sqrt{\Omega_{\rm M}(1+z)^3+\Omega_{\Lambda}}} \, dL_{\gamma} dz,
\label{cumul}
\end{equation}
in which the luminosity--dependent comoving volume neutrino emissivity $j$ is expressed by the product 
of the comoving density of sources with a given $\gamma$--ray luminosity,  provided by the the luminosity 
function $\Sigma(L_{\gamma}, z)$, and the corresponding source neutrino luminosity:
\begin{equation}
j(L_{\gamma}, E_{\nu},z)=\Sigma(L_{\gamma}, z) \, \frac{L_{\nu}({E_{\nu}})}{E_{\nu}}.
\label{emissivity}
\end{equation}
Eq. (\ref{cumul}) is a generalization of the relation used in Paper I, which was suitable 
for a population of sources with a unique luminosity. 

We derive $\Sigma(L_{\gamma}, z)$ using the luminosity function and the parameters for its 
luminosity--dependent  evolution for BL Lacs derived by Ajello et al. (2014) using {\it Fermi}/LAT data.
The local (i.e. $z=0$) luminosity function is described by:
\begin{equation}
\Sigma(L_{\gamma}, z=0)= \frac{A}{\ln(10)L_{\gamma}}\left[ \left( \frac{L_{\gamma}}{L_{\rm *}}\right)^{\gamma_1} 
+ \left( \frac{L_{\gamma}}{L_{\rm *}}\right)^{\gamma_2}   \right]^{-1}
\label{lumfun}
\end{equation}
with $A=3.4\times 10^{-9}$ Mpc$^{-3}$; $\gamma1=0.27$; $\gamma2=1.86$ and $L_{\rm *}=2.8\times 10^{47}$ erg s$^{-1}$.
This luminosity function evolves with $z$ as:
\begin{equation}
\Sigma(L_{\gamma}, z)= \Sigma(L_{\gamma}, z=0) \times e(L_{\gamma}, z),
\end{equation}
where\footnote{The sign of the exponents $p1$ and $p2$ in Ajello et al. (2014) 
is incorrect (M. Ajello, priv. comm.).}:
\begin{equation}
e(L_{\gamma}, z)= \left[ \left( \frac{1+z}{1+z_c(L_{\gamma})}\right)^{-p1(L_{\gamma})} +  
\left( \frac{1+z}{1+z_c(L_{\gamma})}\right)^{-p2}\right]^{-1},
\label{evol}
\end{equation}
and the functions $z_c(L_{\gamma})$ and $p1(L_{\gamma})$ are specified by:
\begin{equation}
z_c(L_{\gamma})=z_c^* \cdot (L_{\gamma}/10^{48} {\rm erg} \, {\rm s}^{-1})^{\alpha},
\end{equation}
\begin{equation}
p1(L_{\gamma})=p1^* +\tau \cdot \log (L_{\gamma}/10^{46} {\rm erg} \, {\rm s}^{-1}).
\end{equation}
The best fit parameters derived by Ajello et al. (2014) are: $p2=-7.4$, $z_c^*=1.34$, 
$\alpha=4.53\times 10^{-2}$, $p1^*=2.24$, $\tau=4.92$.

This parametrization captures the basic features of the $\gamma$--ray emitting BL Lac evolution. 
In particular, low luminosity sources ($L_{\gamma}<10^{45}$ erg s$^{-1}$) are characterized by 
a negative evolution (i.e. a density decreasing with $z$), while sources of higher luminosity display 
a null or positive evolution.

We consider  that the BL Lac $\gamma$--ray luminosity is in the range 
$10^{44}$ erg s$^{-1} < L_{\gamma} < 10^{46}$ erg s$^{-1}$. 
Note that in the {\it Fermi} second AGN catalogue (2LAC, Ackermann et al. 2011) there are sources 
classified as BL Lac objects with even larger $L_{\gamma}$, as also assumed in the population 
study of Ajello et al. (2014). 
However, as discussed in Ghisellini et al. (2011) (see also Giommi et al. 2013 and Ruan et al. 2014), 
these are instead intermediate objects between FSRQ and BL Lacs or even misclassified FSRQ 
whose beamed non--thermal continuum is so luminous to swamp the broad emission lines. 
We suppose that the jets of these sources do not develop an important layer and therefore we 
do not consider them as neutrino emitters.

The assumed maximum luminosity is much below the break luminosity $L_*$. 
The local luminosity function, 
Eq. (\ref{lumfun}), can thus be well approximated by a single power law, 
$\Sigma (L_{\gamma},z=0)\propto L_{\gamma}^{-(\gamma_1+1)}$. 
Recalling the relation between the neutrino and the $\gamma$--ray luminosity ($L_{\nu}\propto L_{\gamma}^2$), 
the neutrino luminosity density (Eq. \ref{emissivity}) can also be expressed as a function of the 
sole $\gamma$--ray luminosity. 
Therefore we can express the contribution of the sources with a given $\gamma$--ray luminosity to 
the total neutrino background as:
\begin{equation}
I(L_{\gamma})\propto L_{\gamma}j(L_{\gamma}) \propto L_{\gamma}\, \Sigma(L_{\gamma}, z) \, 
L_{\nu} \propto L_{\gamma}^{-\gamma_1+2} 
\label{intensity}
\end{equation}
from which $I(L_{\gamma})\propto L_{\gamma}^{1.73}$, i.e. the resulting integral neutrino flux 
is dominated by the most powerful sources.

\section{Results}

%------------------------------------------------------------------
\begin{figure}
\vskip -0.5 cm
\hspace{-1.5 truecm}
\psfig{file=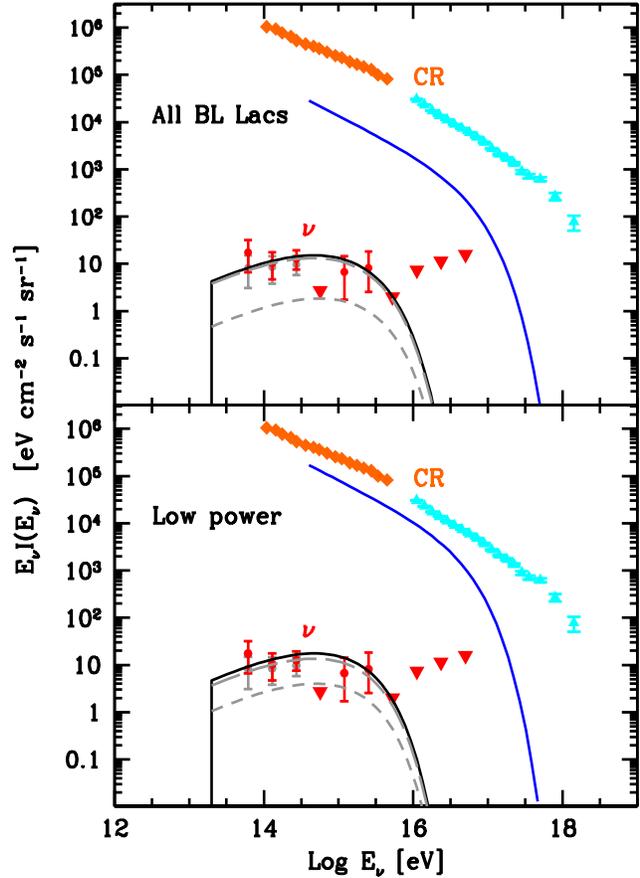,height=12.5cm,width=12.cm}
\vspace{-0.3 cm}
\caption{
{\it Upper panel}: measured diffuse intensities of high--energy neutrinos 
(red symbols, from Aartsen et al. 2014). 
Red triangles indicate upper limits. 
Gray data points show the fluxes for an increase of the prompt atmospheric background to the level of 90\% CL limit. 
The black solid line reports the diffuse neutrino intensity calculated assuming that all the 
BL Lac jets have a spine--layer structure. 
Gray lines report the contributions from sources with 
$10^{44}$ erg s$^{-1} < L_{\gamma} < 10^{45}$ erg s$^{-1}$ (dashed) and 
$10^{45}$ erg s$^{-1} < L_{\gamma} < 10^{46}$ erg s$^{-1}$ (long dashed). 
The blue lines report the corresponding CR intensities, assuming efficient escape from the jet. 
Orange (Apel et al. 2012) and cyan (Chen 2008) data points show the observed high--energy CR spectrum. 
{\it Lower panel:} as the upper panel but considering only the sources with 
$L_{\gamma} <  10^{45}$ erg s$^{-1}$. 
Gray lines show the contribution of sources with 
$10^{44}$ erg s$^{-1} < L_{\gamma} < 3\times10^{44}$ erg s$^{-1}$ (dashed) and 
$3\times 10^{44}$ erg s$^{-1} < L_{\gamma} < 10^{45}$ erg s$^{-1}$ (long dashed).
}
\label{diffuse}
\end{figure}
%------------------------------------------------------------------

\subsection{Cumulative intensity}

We apply the model described above to reproduce the observed neutrino intensity. 
As noted above, the only free parameters of the model are those specifying the proton 
energy distribution and the total CR luminosity normalization: 
$n$, $E^{\prime}_{\rm min}$, $E^{\prime}_{\rm cut}$ and $L^{\prime}_{\rm p,o}$. 
First we consider the case in which the entire BL Lac population is characterized by 
the presence of a structured jet, fixing $L_{\gamma,2}=10^{46}$ erg s$^{-1}$ in Eq. (\ref{cumul}). 
For the parameters reported in the first raw of Tab. 1 we obtain the diffuse spectrum shown by the solid black line in Fig. 2 (upper panel), to be compared with the reported IceCube data points from Aartsen et al. (2014). 
The abrupt cut-off at low energy is an artifact due to the assumed abrupt truncation of the CR energy 
distribution at low energy. 

The gray lines show the contributions from BL Lacs in two different ranges of 
luminosity, namely $10^{44}-10^{45}$ erg s$^{-1}$ (dashed) and $10^{45}-10^{46}$ erg s$^{-1}$ (long dashed). 
As expected from the considerations above (\S 2.3), the total emission is dominated 
by the most luminous sources, although their density is much smaller than that of the low--luminosity ones.

The high--energy cut--off of the CR distribution is robustly fixed to $E^{\prime}_{\rm cut}=3$ PeV 
by the IceCube upper limits at high energy. 
The value of the minimum CR energy $E^{\prime}_{\rm min}$ is instead less constrained. 
The lowest energy IceCube data point at $\approx 100$ TeV allows us to limit $E^{\prime}_{\rm min}$ from above,  
$E^{\prime}_{\rm min}\lesssim 20\times E_{\nu}/\delta_{\rm s}\approx 10^{14}$ eV 
(we ignore the cosmological redshift). 
The curves in Fig. 2 have been derived by assuming $E^{\prime}_{\rm min}=2\times 10^{13}$ eV, 
although lower values are allowed. 
The flat spectrum points to a relatively soft CR spectrum, $n=2.8$.

In the lower panel of Fig. 2 we  report the case, similar to that discussed in Paper I,  
in which only the jets of the weak BL Lacs  --- operatively defined as the sources with 
$L_{\gamma}<10^{45}$ erg s$^{-1}$ --- develop a layer. 
The cosmic ray power for a given $\gamma$-ray luminosity, $L_{\rm p,o}^{\prime}$, increases by a factor 10 with respect to the previous case.

In Fig. 2 we also show the cumulative CR flux from BL Lacs (blue lines) assuming efficient 
escape from the jet and efficient penetration within the Milky Way. 
For the case of all BL Lacs the flux is well below the measured level. 
For the case of HBL alone the flux is close to the limit fixed by the level recorded at the Earth.
This is because, if only low power BL Lacs have to reproduce the neutrino flux,
they must contain a number of energetic cosmic rays which is greater than if BL Lacs of all powers contribute, since, as noted before, the photopion production efficiency $f_{p\gamma}$ -- and hence the neutrino emission efficiency -- increases with the jet power (or, equivalently,  with the $\gamma$-ray luminosity) -- i.e. low power jets are less efficient than high power jets in producing neutrinos.
For our two models the contribution of the BL Lacs to the CR in the $10^{15}-3\times 10^{16}$ eV energy range is of the order of $\sim$5\% and $\sim$50\% for the ``All" and the ``Low power" case, respectively.

% -----------------------------------------------------------
\begin{table} 
\begin{center}
\begin{tabular}{lccc}
\hline
\hline
Model & $L^{\prime}_{\rm p,o}$   & $E^{\prime}_{\rm min}$ &  $E^{\prime}_{\rm cut}$\\
&[erg s$^{-1}$] & [eV] & [eV]\\
\hline  
All& $3\cdot10^{40}$& $2\cdot10^{13}$& $3\cdot10^{15}$ \\
Low power& $3\cdot10^{41}$ & $2\cdot10^{13}$& $2.3\cdot10^{15}$ \\
%Weak HBL & $2.4\cdot10^{43}$& $3\cdot10^{10}$ & $2\cdot10^{15}$ \\
\hline
\hline
\end{tabular}                                                         
\vspace{0.3 cm}
\caption{
Parameters for the two realizations of the model shown in Fig. 2. 
The three columns report the normalization of the CR luminosity, the minimum and the cut--off 
energy of the CR energy distribution.
}
\end{center}
\label{parametri}
\end{table}                                                                  
% --------------------------------------------------------------

\subsection{Jet CR power}

The CR luminosity required to match the observed flux is relatively limited. 
The $\gamma$--ray luminosity dependent beaming--corrected power in CR (similar to that valid for photons, e.g. Celotti \& Ghisellini 2008):

\begin{equation}
P_{\rm CR}\, =\, \frac{L^{\prime}_{\rm p}\, \delta_{\rm s}^4}{\Gamma_{\rm s}^2} \, =\, 
L^{\prime}_{\rm p,o} L_{\gamma,44} \, \frac{\delta_{\rm s}^4}{\Gamma_{\rm s}^2}
\end{equation}
is $P_{\rm CR}\simeq 2\times 10^{43} L_{\gamma,44}$ erg s$^{-1}$ (``All" case) 
and $P_{\rm CR}\simeq 2\times 10^{44} L_{\gamma,44}$ erg s$^{-1}$ (``Low power" case). 
This value can be compared to the beaming corrected radiative luminosity, which for blazars 
can be directly related to the observed $\gamma$--ray luminosity (Sbarrato et al. 2012), 
$P_{\rm rad}\approx 3\times 10^{42} L_{\gamma,44}^{0.78}$ erg s$^{-1}$.
The ($\gamma$--ray luminosity dependent) ratio between the two quantities is 
thus $\xi=P_{\rm CR}/P_{\rm rad}\approx 5 \, L_{\gamma,44}^{0.22}$ and 
$\approx 50 \, L_{\gamma,44}^{0.22}$ for the two cases. 
These values should be compared to
$\xi\approx 100$ assumed by Murase et al. (2014) -- although the possible existence of a curved CR distribution as that discussed by Dermer et al. (2014) should allow to reduce such a large value.

Since for blazar jets the ratio between the radiative and the kinetic power 
(calculated assuming a composition of one {\it cold} proton per emitting electron) is 
$P_{\rm rad}/P_{\rm jet}\approx 0.1$ (e.g. Nemmen et al. 2012, Ghisellini et al. 2014), 
we can also assess the ratio between the jet power
(calculated neglecting the contribution of CR) to the CR power, 
$P_{\rm CR}/P_{\rm jet}\approx 0.5 \, L_{\gamma,44}^{0.22}$ for the ``All" case and 
ten times larger for the ``Low power" case. Therefore, even in the most conservative case, 
the jet should be able to channel a sizable part of its kinetic power into CR acceleration.
As a consequence, the total jet power should increase by a corresponding amount 
with respect to the current estimates.

\subsection{Neutrino point sources}

Having calculated the expected cumulative neutrino flux we can also derive the expected number of 
events detectable by IceCube from a given BL Lac object. 
This is particularly valuable in view of the identification of the possible astrophysical counterparts 
of the detected neutrinos and to test our model.

To this aim, first of all we calculate the theoretical differential neutrino number flux at the Earth 
from a generic source of neutrino luminosity $L_{\nu}$ at redshift $z$ as:
\begin{equation}
\phi(E_{\nu})\equiv\frac{dN}{dt\, dE_{\nu}\, dA}=\frac{L_{\nu}[E_{\nu}(1+z)]}{4\pi d_L^2 \, E_{\nu}},
\end{equation}
where $d_{\rm L}$ is the luminosity distance. 
We derive the expected IceCube rate convolving the flux with the energy--dependent IceCube 
effective area $A_{\rm eff}$ (taken from Aartsen et al. 2013a). 
Finally we derive the number of events expected with an exposure of 3 years (corresponding 
to an effective exposure of $T_{\rm exp}=998$ days):
\begin{equation}
N_{\nu}=T_{\rm exp} \int A_{\rm eff}(E_{\nu}) \phi(E_{\nu}) \,dE_{\nu}.
\end{equation}
For consistency, we first checked that our two models represented in Fig. 2 provide about 
30 events detected in 3 years (Aartsen et al. 2014).
We then applied the procedure using the BL Lacs belonging to the 2LAC 
catalogue\footnote{{\tt http://www.asdc.asi.it/fermi2lac/}} (Ackermann et al. 2011) with measured redshift. 
For each source the $\gamma$--ray luminosity is derived converting the flux provided by the 
2LAC catalogue using the procedure described in Ghisellini et al. (2009). 
The $\gamma$--ray luminosity is then converted into the luminosity in neutrinos 
$L_{\nu}(E_{\nu})$ according to our model. 
The resulting event rate ($N_{\nu}/T_{\rm exp}$ in units of yrs$^{-1}$) for the sources, 
as a function of $L_{\gamma}$ are reported in Fig. 3 for the two scenarios. 
The number of events for the exposure of 998 days $N_{\nu}$ for the five brightest sources 
and for the two possible scenarios explored above are reported in Table 2. 

% ----------------------------------------------------------------
\begin{figure}
%\hskip -2 cm
\hspace{-1.5 truecm}
%\vspace*{-1 truecm}
%\vskip -0.3 cm
\psfig{file=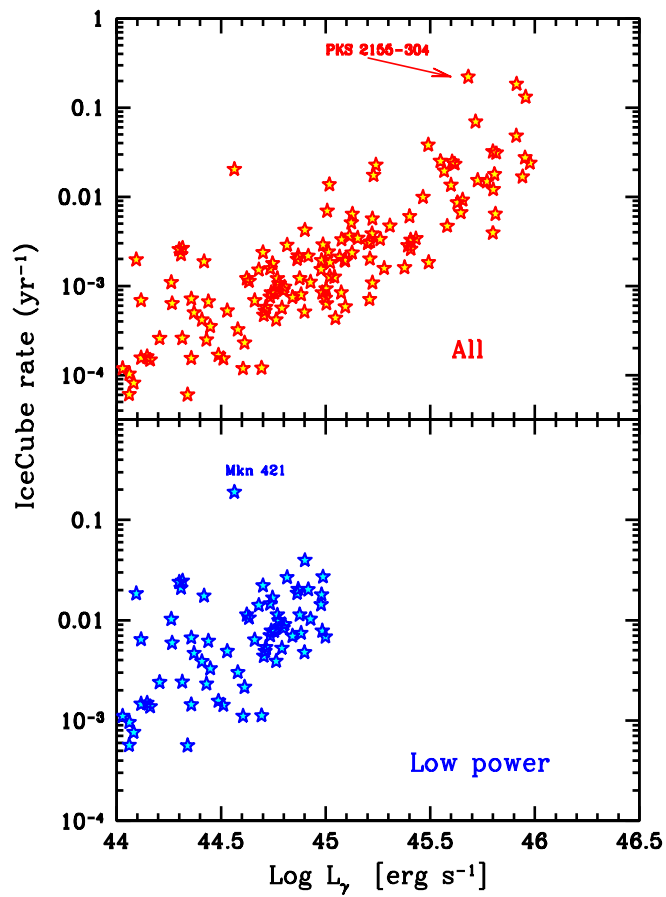,height=12.5cm,width=12.cm}
\vspace{-0.1 cm}
\caption{
Expected IceCube count rate (events/year) for the 2LAC BL Lac as a function of the $\gamma$--ray 
luminosity for the ``All" (upper panel) and the ``low power" (lower panel) scenario, respectively. 
}
\label{diffuse}
\end{figure}
% ----------------------------------------------------------------

For the ``All" scenario the brightest 3--4 sources are characterized by a rate sufficient to allow 
the detection of several events with a relatively prolonged exposure. 
The most probable candidate is PKS 2155--304 (see below). 
On the other hand, if only low power BL Lacs are considered, the situation is quite different, with 
only one source --- Mkn 421 --- expected to be detectable and with all the other sources with a 
rather smaller flux, providing $N_{\nu}<0.1$. 
Note also that both sources, with an expected number of events in 3 years $N_{\nu}\approx1$ 
(over a total of about 30 neutrino detected) should be characterized by a neutrino flux of 
the order of $E_{\nu}^2\phi(E_{\nu})\approx 4\pi E_{\nu} I(E_{\nu}) /30\approx 10^{-11}$ erg cm$^{-2}$ s$^{-1}$.

An obvious {\it caveat} is in order when considering this result. 
This calculation, although applied to single sources, is built on our results based on the 
{\it averaged} characteristics of the BL Lac population (besides our model assumptions). 
Furthermore, the $\gamma$--ray luminosity of the 2LAC is an average over 2 years of observations. 
Given these limits, our procedure cannot consider source peculiarities or mid--term variability, 
particularly relevant for the high--energy emission of luminous BL Lac objects (e.g., Abdo et al. 2010).

% -----------------------------------------------------------
\begin{table} 
\begin{center}
\begin{tabular}{lcc}
\hline
\hline
Source & $z$ & $N_{\nu}$ (998 days)\\
\hline  
\multicolumn{3}{c}{All}\\
\hline
PKS 2155-304& 0.116& 0.6\\
PKS 0447-439& 0.205& 0.5\\
PKS 0301-243& 0.26  & 0.4\\
1H 1013+498  & 0.212& 0.2\\
S4 0954+65    &0.367 &  0.15\\
\hline  
\multicolumn{3}{c}{Low power}\\
\hline
Mkn 421& 0.031 & 0.5\\
1ES 0806-05& 0.137& 0.1\\
RX J 0159.5+1047& 0.195& 0.1\\
1ES 1959+650& 0.047& 0.1\\
1ES 2322-409& 0.062& 0.05\\
\hline
\hline
\end{tabular}                                                         
\vspace{0.3 cm}
\caption{
List of the BL Lacs and the expected neutrino counts for an exposure of 3 years and the 
two scenarios described in the text.
}
\end{center}
\label{events1}
\end{table}                                                                  
% --------------------------------------------------------------

\begin{figure}
%\hskip -2 cm
\vspace*{-1.2 truecm}
\hspace*{-0.3 truecm}
%\vspace*{-1 truecm}
%\vskip -0.3 cm
\psfig{file=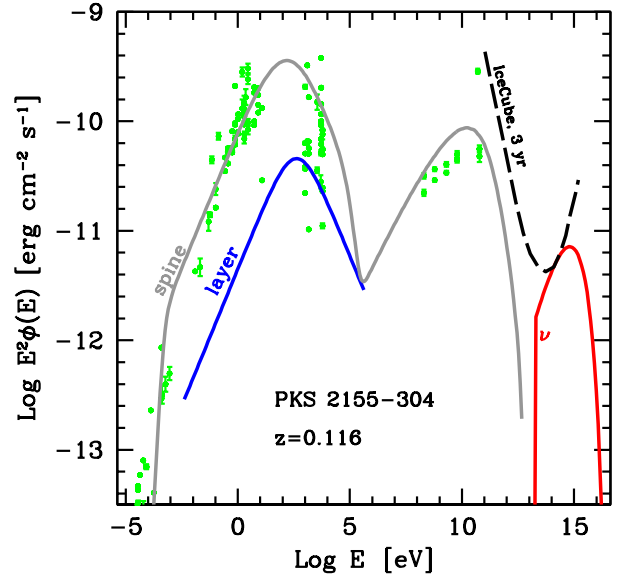,height=10.cm,width=10.cm}
\vspace{-1.5 cm}
\caption{
Spectral energy distribution of PKS 2155--304 (green points --- taken from {\tt http://www.asdc.asi.it} --- 
and solid gray line). 
The red solid line shows the expected neutrino emission expected in our model for the "All BL Lac" case. 
The blue line tracks the corresponding layer emission. 
The black dashed line marked ``IceCube, 3 yr" displays the estimated flux limit for IceCube, obtained scaling the 
sensitivity curve provided in Tchernin et al. (2013).
}
\label{sed}
\end{figure}

Given these limitations, it is however possible to note a clear difference between the two cases: 
in the "All" case there is a bunch of relatively bright neutrino BL Lacs -- those with the highest power. 
For the ``Low power" case, instead, Mkn 421 largely dominates over the other sources.
A remark concerns the case PKS 2155--304, the first entry in Table 2 for the ``All" case, 
which is the only highly peaked BL Lac (HBL) of this list. Indeed, as we noted above, 
the neutrinos reaching the Earth are preferentially produced by the most powerful sources 
which preferentially are of the intermediate (IBL) or low peaked (LBL) type (e.g., 
Ackermann et al. 2011, Giommi et al. 2012). 
PKS 2155--304 is clearly an outlier of this general trend, displaying a SED typical of HBL 
(i.e. the synchrotron peak in the soft X-ray band) but with a luminosity much larger than that 
of the averaged HBL population. 
Given the link that we assumed between the electromagnetic and neutrino output, this peculiarity 
shows up also in the neutrino window. 
In Fig. \ref{sed} 
we report in detail the SED and the expected neutrino output for this source. 
In the SED we also report the IceCube sensitivity curve for 3 years, scaling that provided 
in Tchernin et al. (2013) for one year and about half of the detector (IC-40 configuration). 
Clearly, the flux limit is very close to that theoretically expected. 

Recently, Ahlers \& Halzen (2014) performed a calculation aimed to assess the possibility 
to single--out neutrino sources with IceCube, considering different possible source scenarios 
and taking into account the details of the the IceCube instrument (e.g. background, statistics). 
They also consider blazars as possible sources, adopting the local density and the cosmological 
evolution valid for the most powerful blazars, i.e. FSRQ. Comparing the neutrino flux they 
derived for single sources with the upper limits on the flux of the the weak BL Lac Mkn 421 and Mkn 501, 
they conclude that a blazar origin is disfavored. 
However, as said, their calculations are clearly tuned for FSRQ, not BL Lacs, which display 
quite different cosmological density and evolution. Indeed, a  calculation based on the 
BL Lac demography provides results compatible to those presented here (M. Ahalers, priv. comm.).

\section{Discussion}

We have presented an extension of the scenario envisaging the production of high--energy neutrinos 
in the structured jets of BL Lac objects sketched in Tavecchio et al. (2014). 
The key ingredient is the relativistic boosting of the radiation produced in the layer in the 
spine frame (Ghisellini et al. 2005, Tavecchio \& Ghisellini 2008), which entails the increased 
efficiency of the photo--pion reactions and the following neutrino emission.

% --------------------------------------------------------------
% --------------------------------------------------------------

The observational evidence supporting the idea that BL Lac (and radiogalaxy) jets are structured outflows  
--- i.e. with a faster spine surrounded by a slower layer --- is steadily accumulating. 
The deceleration of the flow after the blazar region expected for this configuration offers the simplest 
explanation of the anomalous lower apparent speeds inferred for jets of TeV emitting BL Lac through 
VLBI observation (Piner \& Edwards 2004, Piner et al. 2008). 
Likely, a spine--layer system is the most natural description of the limb brightening displayed 
by several BL Lac (Giroletti et al. 2004, 2008, Piner \& Edwards 2014, Piner et al. 2010) and 
radiogalaxy (e.g., Nagai et al. 2014, M{\"u}ller et al. 2014) jets at VLBI resolutions. 
Strong independent --- although indirect --- support is provided by arguments from to the 
unification scheme of BL Lacs and Fanaroff-Riley I (FRI) radiogalaxies (e.g. Chiaberge et al. 2000, Meyer et al. 
2011, Sbarrato et al. 2014). Indeed, while large Lorentz factors ($\Gamma\approx 10-20$) are required 
to model the BL Lac emission (e.g. Tavecchio et al. 2010), the emission properties and the number 
densitiy of radiogalaxies instead favor low ($\Gamma\approx 3-5$) bulk Lorentz factors. 
The structure jet scenario easily solves this problem: depending on the jet viewing angle, large 
or low Lorentz factors are inferred for BL Lac (dominated by the spine) or radiogalaxies (for 
which the layer contributes most to the emission), respectively.

There is relatively small number of cases for which some constraints to the structural parameters of 
the layer can be derived (e.g., Tavecchio \& Ghisellini  2008, 2014). 
In our work we assumed a phenomenological view, tuning the layer properties (bulk Lorentz factor, 
emitted spectrum) so that we can reproduce at best the observed neutrino flux. 
An improvement of present knowledge could help in better test our proposal.

One is naturally led to wonder whether also misaligned BL Lacs -- i.e. FRI radiogalaxies, according to the classical unification scheme for radio-loud AGN (e.g. Urry \& Padovani 1995) -- could contribute to the observed neutrino background (see also Becker Tjus et al. 2014). In this case, one expects that the close-by radiogalaxies Cen A,  M87 and NGC 1275 -- also observed to emit TeV photons (Aharonian et al. 2009, Aharonian et al. 2006, Aleksic et al. 2012) --  should be optimal candidates for a direct association with IceCube events (recall also that Cen A is also possibly associated to a handful of UHECR detected by AUGER, Abraham et al. 2007). For all three sources, however, quite stringent upper limits are derived (Aartsen et al. 2013b). In the structured jets scenario adopted here, the electromagnetic emission from the inner jet of radiogalaxies (at least at high energies, see below), is likely to be dominated by the layer, since, due to the large viewing angle, the more beamed spine radiation is de-boosted as observed from the Earth. Analogously, we expect that possible neutrino  emission from the layer would dominate over that of the spine in case of misaligned jets. As for the inverse Compton emission (e.g. Tavecchio \& Ghisellini 2008), the dominant radiation field for the photo-pion reaction is expected to be that of the spine, boosted in the layer frame by the relative motion. For M87 and NGC 1275, the application of the spine-layer scenario (Tavecchio \& Ghisellini 2008, 2014) suggests that the high-energy $\gamma$-ray component is produced in the layer, while the  low energy non-thermal emission 
is rather due to the (de-boosted) spine emission. In this case we therefore have a direct handling on the spine radiation field. For all the aforementioned radiogalaxies this low energy components peaks in the IR band, around $\epsilon_{\rm o}\approx 0.1$ eV. Therefore, in order to allow the photo-meson reaction, proton energies  should exceeds (we neglect the small Doppler shift) $E_{\rm p}\gtrsim m_{\pi}m_{\rm p}c^4/\epsilon_{\rm o}\approx 10^{18} (\epsilon _{\rm o}/{\rm 0.1 \,\, eV})^{-1}$ eV, implying that the resulting neutrinos have energies exceeding $E_{\nu}\gtrsim 50$ PeV, well above the energies of neutrinos considered here. If such high energy for protons are attainable, upcoming new detectors extending beyond the IceCube band (ARA, Allison et al. 2012; ARIANNA, Barwick 2007; ANITA, Gorham et al. 2009; EVA, Gorham et al. 2011) could thus be able to detect neutrinos from the layer of nearby radiogalaxies.

An attractive feature of our scenario, especially in the case for which the entire BL Lac population 
contributes to the observed flux, is the moderate required power in CR. Indeed, while it is 
typically assumed that the CR luminosity greatly exceeds that in radiation (e.g. Murase et al. 2006, 2014), 
we found that a ratio $P_{\rm CR}/P_{\rm rad}\approx 5$ can match the observed flux. 
In turn, the ratio between the CR power and the kinetic power --- as derived through the modeling of 
the of the observed emission with  standard leptonic models, e.g. Ghisellini et al. (2014) --- is 
$P_{\rm CR}/P_{\rm jet}\lesssim 1$. 
In case of fast cooling of the accelerated electrons, $P_{\rm CR}/P_{\rm rad}$ directly provides a 
measure of the electron--to--proton luminosity ratio, $f_{\rm e}\approx P_{\rm rad}/P_{\rm CR}$.  
Theoretical expectations indicate values $f_{\rm e}\ll 1$, (e.g., Becker Tjus et al. 2014 
and references therein), consistent with our findings.  
The fact that the CR power can be a sizable fraction of the jet power could perhaps be linked 
to the deceleration of the jet from sub--pc to pc scale as inferred from VLBI observations 
(e.g., Piner \& Edwards 2014). We also recall that propagating CR beams produced by BL Lac jets have been invoked to explain several peculiarities of low-power BL Lac jets (the so-called extreme HBL), in particular their hard and slowly variable TeV emission (e.g. Essay et al. 2010, Murase et al. 2012, Aharonian et al. 2013, Tavecchio 2014)

It should be remarked that the CR power sensitively depends on the minimum energy. 
In our modeling we assumed $E^{\prime}_{\rm min}\sim 10^{13}$ eV, as limited by the observed 
low energy data points. 
Lower values are not excluded, of course, possibly increasing the energy budget. 
A related point concerns the required maximum CR energy. 
The IceCube upper limits robustly fixed the (spine rest frame) maximum energy to few PeV. 
General considerations (e.g. Tavecchio 2014) allow us to estimate that the maximum energy 
of the accelerated protons to be $E_{\rm max}\simeq 3\times 10^{17} R_{15} B/\epsilon$, 
where $\epsilon>1$ is a parameter, incorporating the details of the acceleration mechanism, 
determining the acceleration efficiency and $B$ is  the magnetic field. 
For BL Lac jets typical values are $B=0.1-1$ G (e.g., Tavecchio et al. 2010).
Energies of the order of few PeV could thus be reproduced for $\epsilon \approx 0.01-0.1$.

We provided a list of the sources with the largest expected neutrino flux, which are the best candidates to be 
detected as point-sources by IceCube. 
The kind and the characteristics of the sources are quite different in the two scenarios. 
In the case in which neutrino emission occurs in all the BL Lac population the brightest sources are those 
with powerful jets (IBL and LBL type) located at relatively large redshift ($z\sim 0.2$). 
The most probable source is PKS 2155--304, a HBL with an atypically large luminosity. 
In the case in which, instead, only low power jets have an efficient layer, the most probable sources 
associated to neutrino events are HBL at low redshift. 
In both cases we expect that the brightest sources could have several associated neutrinos in the next few years. If BL Lac objects are the sources dominating the extragalactic neutrino sky, our model can thus be 
effectively tested by a more extended IceCube exposure and the two options that we 
presented could be effectively distinguished.

\section*{Acknowledgments}
FT acknowledges  contribution from a grant PRIN--INAF--2011. 
Part of this work is based on archival data and  on--line services 
provided by the ASI Science Data Center.

\end{document}